\documentclass[conference]{IEEEtran}
\IEEEoverridecommandlockouts
\usepackage{cite}
\usepackage{marvosym}
\usepackage{amsmath,amssymb,amsfonts}
\usepackage{graphicx}
\usepackage{bm}
\usepackage{balance} 

\usepackage[noend]{algorithmic}

\usepackage{url}
\usepackage{algorithm}
\usepackage{algorithmic}
\usepackage{amsmath}
\usepackage{amssymb}

\usepackage{subfigure}
\usepackage{textcomp}
\usepackage{xcolor}
\usepackage{color, soul}

\newtheorem{theorem}{Theorem}

\def\BibTeX{{\rm B\kern-.05em{\sc i\kern-.025em b}\kern-.08em
    T\kern-.1667em\lower.7ex\hbox{E}\kern-.125emX}}
\begin{document}
 
 \sethlcolor{yellow} 

\title{3D Wireless Channel Modeling for Multi-layer Network on Chip}
\author{\IEEEauthorblockN{1\textsuperscript{st} Chao Ren}
\IEEEauthorblockA{\textit{School of Computer and Communication} \\
\textit{Engineering and Shunde Graduate School,} \\
\textit{University of Science and Technology Beijing} \\
Beijing, China\\
chaoren@ustb.edu.cn}
\and
\IEEEauthorblockN{2\textsuperscript{st} Jingze Hou}
\IEEEauthorblockA{\textit{University of Science} \\
\textit{and Technology Beijing} \\
Beijing, China \\
jingzehou@163.com}
\and
\IEEEauthorblockN{3\textsuperscript{rd} Biao Pan\textsuperscript{\Letter}}
\textit{MIIT Key Laboratory of}\\
\textit{Spintronics, School of Integrated}\\
\textit{Circuit Science and Engineering}, \IEEEauthorblockA{\textit{Beihang University } \\
Beijing, China \\
panbiao@buaa.edu.cn}
}

\maketitle

\begin{abstract}
 The resource constraints and accuracy requirements for Internet of Things (IoT) memory chips need  three-dimensional (3D) monolithic integrated circuits, of which the increasing stack layers (currently more than 176) also cause excessive energy consumption and increasing wire length.
In this paper, a novel 3D wireless network on chips (3DWiNoCs) model transmitting signal directly to the destination in arbitrary layer is proposed and characterized. 
However, due to the the reflection and refraction characteristics in each layer, the complex and diverse wireless paths in 3DWiNoC add great difficulty to the channel characterization.
To facilitate the modeling in  massive layer NoC situation, both boundary-less model  boundary-constrained 3DWiNoC model are proposed, of which the channel gain can be obtained by a computational efficient  approximate algorithm.
These 3DWiNoC models with approximation algorithm can well characterize the 3DWiNoC channel in aspect of complete reflection and refraction characteristics, and avoid massive wired connections, high power consumption of cross-layer communication and high-complexity of 3DWiNoC channel characterization. 
Numerical results show that:
1) The difference rate between the two models is lower than $0.001\%$ (signal transmit through 20 layers);  
2) the channel gain  decreases sharply if refract time increases;
and 3) the approximate algorithm can achieve an acceptable accuracy (error rate lower than $0.1\%$).

\end{abstract}


\section{Introduction}

In recent years, the number of cores and threads in chips of mobile devices has been increasing to meet the requirement of Internet of Things (IoT). To reduce the area overhead, 3D integration technology are widely used in IoT memory devices
  \cite{5306555}.
 However, the increase of through-silicon-vias (TSV) density in 3D chips leads to higher wiring complexity as well as higher cost of wired communication between cores \cite{4550025}. 
Even though the transmitter and receiver are close to each other, the length of the wire increases due to layers integration in 3D chips.
 In addition, the greatly extended length of wire causes high latency and energy consumption which hinder the deployment of these 3D chips on IoT platforms. 
To directly set up the link between two nodes and avoid the tedious path-selection processing in 3D integrated circus, wireless network on chip is considered a promising approach \cite{6991560}. 
 
For wireless network-on-chips (WiNoCs), the plane wave is used as the main signal carrier rather than the space wave \cite{959649}.
Ref. \cite{5551123} and \cite{4339598} further demonstrated that at least three types of plane waves on which the signal is propagated over an intra-chip channel must be considered: space waves (air), surface waves (air-wafer interface), and guided waves (through a silicon substrate).
Likewise, these works conclude that the dominant path is transmitted by surface waves, which means the main path in WiNoCs is non Line-of-Sight (relying on refraction and reflection).
In Ref. \cite{6525613},  path loss and dispersion are shown as additional channel-influencing factors of WiNoC, and dispersion effect can be avoid in practice by using certain frequency. 
The authors in \cite{8406954} and \cite{5067340} characterized the wireless channel of single-layer WiNoC by integrating the signals from all angles and classifying the times of reflection and refraction in different materials.
However, these analyses are based on 2D WiNoCs and not able to fully characterize the wireless 3D WiNoC channel.
The rays reflecting and refracting between any two layers in 3D WiNoC leads to great  complexity and diverse of massive paths, which shows great difficulty in characterizing the wireless channel.
To the best of our knowledge, channel modeling of WiNoCs with multiple layers has not been well studied in previous works.

In this paper, novel channel modeling for 3D  wireless network on chip (3DWiNoC) is studied, which fully considers the reflect and refract paths between multiple layers.
The modeling of 3DWiNoC channel is started with analyzing the  reflection and refraction times of a single path, and then extends to more paths separately.
The modeling can be processed in four steps:
1) Determine the range of transmitting angles of the signal that can be received;
2) Analyze each path's reflection and refraction between any two materials in the stack;
3) Classify the paths according to the number of reflect and refract paths;
4) Obtain the gain of each path and thus calculate the total channel gain.

As a means of simplification, the practical boundary-constrained 3DWiNoC model can be idealized into boundary-less one.
This paper shows that compared with the boundary-constrained model, the boundary-less one is much simple (with lower computation loops, and the proof is shown in \textbf{Appendix A}) and the gap between simulation result is minor (please refer to Fig. 13).
Thus, to quickly get an approximated result, we can replace the boundary-constrained model with the boundary-less one, especially when the transmitter is far away from the boundary.

To further reduce the complexity of calculating the paths introduced by reflection and refraction in 3DWiNoCs, an approximation algorithm excluding the path with an exceeding small gain and large delay is proposed.
This algorithm significantly reduces computation loops while its accuracy is acceptable (error rate is lower than $0.1\%$)

The contributions of this paper are the following.
\begin{itemize}
\item  A novel wireless channel model, named 3DWiNoC, is proposed for multi-layer network on chip, which fully considers reflection  and  refraction effects.
\item  A computational effective approximation algorithm is proposed to obtain 3DWiNoC channel gain with acceptable accuracy. 
  
\end{itemize}
This rest of this paper is organized as following. In Section II, the characteristics of wireless 3D NoCs are modeled. In Section III, the proposed path-classification algorithm is presented. Section IV provides numerical results and Section V concludes the paper.

\section{System Model}

\begin{figure}[t]
\center
\centering\includegraphics[scale=0.58]{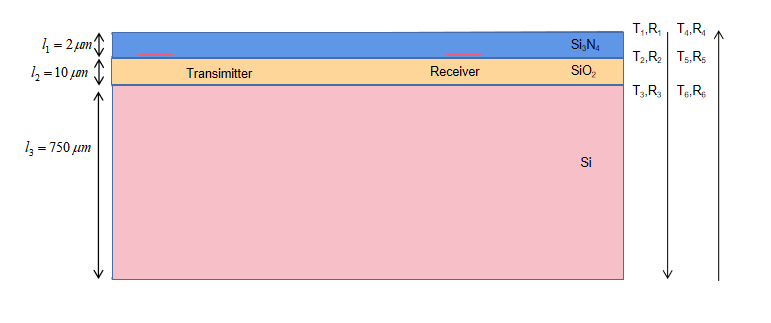}
\caption{Model of single-layer NoCs. (The transmitter and receiver are marked with red color.)}
\label{fig_modelsingle}
\end{figure}

\begin{figure}[t]
\center
\centering\includegraphics[scale=0.6]{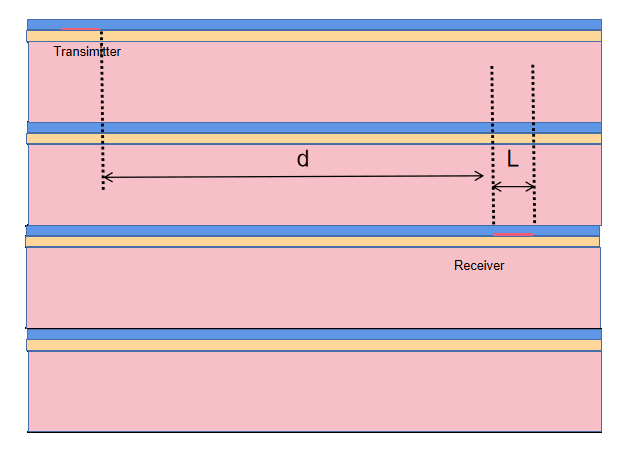}
\caption{Model of 3DWiNoCs.}
\label{fig_modelmulti}
\end{figure}

3DWiNoCs consists of multiple single layer WiNoCs (170 layers or more).
Each layer of the NoCs is made up of three material stacks, and the thickness of material stacks are  $l_{1}$, $l_{2}$, and $l_{3}$ respectively.
For single layer model, the transmitter and receiver are located at the bottom of the $Si_{3}N_{4}$ layer (see Fig. \ref{fig_modelsingle}).
The transmitter and receiver are the antenna in package (AiP) to meet the narrow space constraint in recent chips.
The substrate (typically consists of multiple layers) under the antennas can sharply reduce the gain of signal.
As a result, we use the condition that receiver receives the signal from the upper layer (see Fig. \ref{fig_modelmulti}).
The transmitter and receiver are shown in Fig. \ref{fig_modelmulti} as $T$ and $R$, respectively. 

 In Fig. \ref{fig_modelmulti}, the number of layers between the transmitter and the receiver is \textit{J}.
 The antenna length of the receiver is \textit{L}, and the horizontal displacement between the transmitter and receiver is \textit{d}.
 $T_{1}$, $T_{2}$, $T_{3}$, $T_{4}$, $T_{5}$ and $T_{6}$ are the refraction coefficients of a wave from one medium to another, and $R_{1}$, $R_{2}$, $R_{3}$, $R_{4}$, $R_{5}$, and $R_{6}$ are the reflection coefficients of a wave from one medium to another (see Fig. \ref{fig_modelsingle}).
 The relationship between refraction coefficients and reflection coefficients can be formulated as (\ref{equ:tr}).
 
 \begin{equation}
\left\{
 \begin{array}{lr}
  T_{1}=\frac{(n_{3}-n_{1})^{2}}{(n_{3}+n_{1})^{2}} &  \\
  T_{2}=\frac{(n_{2}-n_{1})^{2}}{(n_{2}+n_{1})^{2}} & \\
  T_{3}=\frac{(n_{3}-n_{2})^{2}}{(n_{3}+n_{2})^{2}} &  \\
  T_{4}=T_{1}    &  \\
  T_{5}=T_{2}    &  \\
  T_{6}=T_{3}    &  \\
  R_{i}=1-T_{i}, &  i=1,2,\cdots,6
             
             \end{array}
\right.
\label{equ:tr}
\end{equation}

 The paths that travel through the top layer can be neglected because the oxide thickness is negligibly small compared to a wavelength.
 Thus, the signal will only be reflected between $Si$ layer and the other material layers.
 
 \begin{figure}[t]
\center
\centering\includegraphics[scale=0.45]{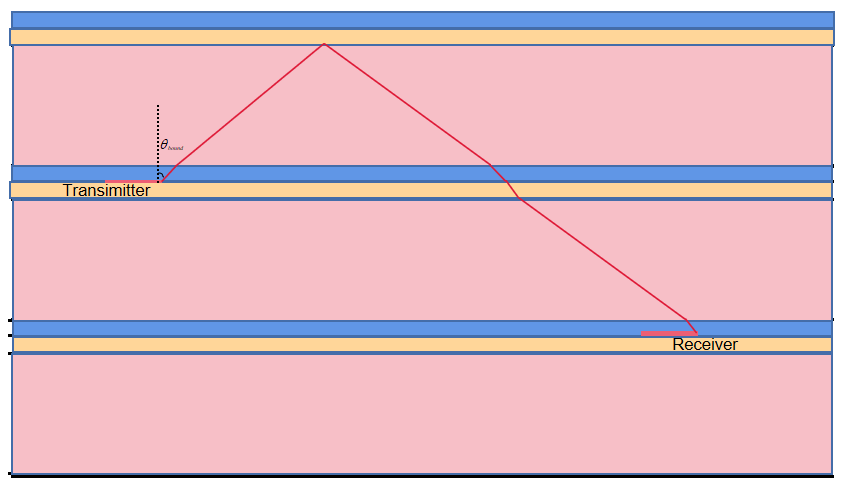}
\caption{Path with transmitting angle $\theta_{bound}$.}
\label{fig_modelpath1}
\end{figure}

Considering that the complete refraction and reflection process of the path in 3DWiNoC is complex and difficult to completely and clearly described, the modeling can be implemented in the following order (1-3).

(1) Analyze the transmission angle based on the reachability of paths from the transmitter to the receiver.

(2) Analyze the possible refraction step time of each path after the transmission angle is determined.

(3) Analyze possible reflection step time of the path after the transmission angle and the time of refraction step is determined.

Because only a fraction of transmitted rays from transmitting angle set can be received, a critical boundary of angle ($\theta_{bound}$) exists.
When the transmitting angle is larger than $\theta_{bound}$, the corresponding path can not reach the receiver (see Fig. \ref{fig_modelpath1}).

\begin{equation}
\begin{aligned}
 (d+L)=&\left(l_{1}\tan{\arcsin{\frac{n_{1}\sin{\theta}} {n_{3}}}}+l_{2}\tan{\arcsin{\frac{n_{2}\sin{\theta}} {n_{3}}}}\right)\times\\
 (2q+J)+&(2q+J+2)l_{2}\tan{\theta}+2l_{1}\tan{\arcsin{\frac{n_{1}\sin{\theta_{bound}}} {n_{3}}}}
	\label{equ:thetabound}
\end{aligned}.
\end{equation}

Based on (\ref{equ:thetabound}), we can get the critical angle $\theta_{bound}$ to determine what path can reach to the receiver.

After determining the angle at which the signal can reach the receiver, it is necessary to determine the characteristics of the path to the receiver (including the number of reflect and refract rays involved).
The horizontal displacement of all received paths is within the length range of the receiver.

 \begin{figure}[t]
\center
\centering\includegraphics[scale=0.6]{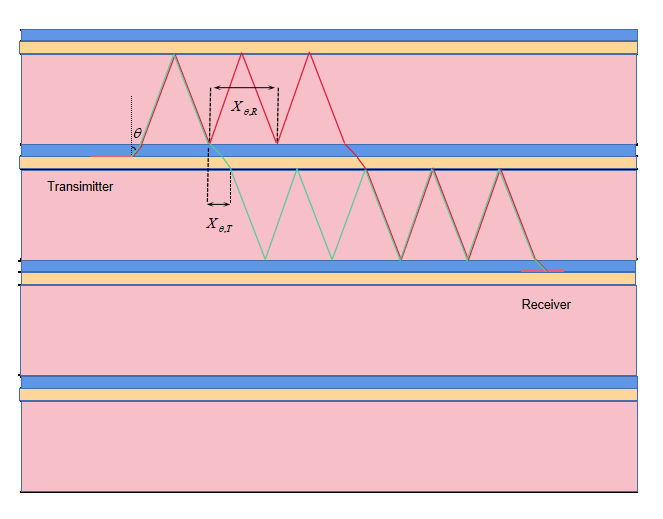}
\caption{Paths in 3D WiNoCs.}
\label{fig_modelpath}
\end{figure}

For the path in a given transmitting angle $\theta$, the horizontal displacement variables for each reflection step (the path reflects twice in $Si$) and refraction step (the path travels through two materials) are defined as $X_{\theta,R}$ and $X_{\theta,T}$ respectively (see (3), (\ref{equ:xtr}) and Fig.  \ref{fig_modelpath}).

\begin{equation}
	X_{\theta,T}= l_{1}\tan{\arcsin{\frac{n_{1}\sin{\theta}} {n_{3}}}}+l_{2}\tan{\arcsin{\frac{n_{2}\sin{\theta}} {n_{3}}}}
	\label{equ:xtt}
\end{equation}

\begin{equation}
	X_{\theta,R}= 2l_{2}\tan{\theta},
	\label{equ:xtr}
\end{equation}
where $n_{1}$, $n_{2}$, and $n_{3}$ are the indexes of refract rays of a signal in each material layer.

Since the receiving antenna has a certain length, any signal within the receiving range can be received.
The signal has been refracted $tra_{\theta}$ times before reaching the receiver at a certain $\theta$. Due to the number of refraction step is integer and continuously changing, $tra_{\theta}$ is a arithmetic progression with a tolerance of 1.

Since the transmitter is $J$ layers away from the receiver, the signal reaches the receiver for refracting at least $J$ times:

\begin{equation}
	tra_{\theta}[1]= J
	\label{1}
\end{equation}

When the time of refraction is maximum, the path satisfies the condition that it is near the right endpoint of receiver (the displacement of the path is maximum) and consistent with the refraction and the necessary reflection effects (reflection rays is needed to ensure the continuous of refraction). (see Fig. \ref{fig_modelpath1}) with the most time of refraction, $tra_{\theta}[x_{\theta}]$ is:
 
 \begin{equation}
	tra_{\theta}[x_{\theta}]= \left\lfloor \frac{2(d+L-2l_{1}\tan{\arcsin{\frac{n_{1}\sin{\theta}} {n_{3}}}}-X_{\theta,R})}{2X_{\theta,T}+X_{\theta,R}} \right\rfloor
	\label{equ:tra}
\end{equation}

\begin{equation}
	x_{\theta}= tra_{\theta}[x_{\theta}]-tra_{\theta}[1]+1,
	\label{equ:xtheta}
\end{equation}
where $x_{\theta}$ is a variable related to the transmission angle $\theta$ and $tra_{\theta}[x_{\theta}]$ describes the maximum number of refraction step of the path (the path with transmission angle $\theta$).

Baesd on (\ref{1}) and (\ref{equ:tra}), all possible refracted times constitute a matrix $\textbf{Tra}_{\theta}$.
\begin{align}
	\textbf{Tra}_{\theta}&=\left [
	\begin{matrix}
		tra_{\theta}[1]\\
		tra_{\theta}[2]\\
	    \vdots\\
		tra_{\theta}[x_{\theta}]\\
	\end{matrix}\right ],
\end{align}
where $tra_{\theta}[x_{\theta}]$ and $x_{\theta}$ are defined in (\ref{equ:tra}) and (\ref{equ:xtheta}).

Next we should analyse the possible reflection time of path after  $\theta$ and $n_{\theta}$ (time of refraction step) are determined.
The signal can be received when its horizontal displacement of reflection and refraction added up is between $d$ (the left endpoint of the receiver) and $d+L$ (the right endpoint of the receiver).
It is assumed that the path is reflected $m_{\theta}$ times, and (\ref{relation}) give the premise condition of the  path reaching to the receiver.

\begin{equation}
\begin{aligned}
	d\leq n_{\theta}X_{\theta,T}
	+(m_{\theta}+\frac{n_{\theta}}{2}+1)X_{\theta,R}+&
	\\\l_{1}\tan{\arcsin{\frac{n_{1}\sin{\theta}} {n_{3}}}} &\leq d+L.
	\label{relation}
	\end{aligned}
\end{equation}

For each determined $n_{\theta}$ calculated by (5), (6) and (9), there is a maximum and a minimum number for reflect rays.
When the path approaches to the left endpoint of the receiver after $n_{\theta}$ times refraction, the time of reflection is minimum (shown as (\ref{ref1})).
When the path approaches to the right endpoint of the receiver after $n_{\theta}$ times refraction, the time of reflection is maximum (shown as (\ref{equ:refth})).
Because the time of reflection is changing continuous, $ref_{\theta,n_{\theta}}$ is a arithmetic progression with tolerance of 1, so the number of the possible reflection is $y_{(\theta,n_{\theta})}$ (shown as (\ref{equ:ytheta})).

\begin{equation}
\begin{aligned}
	ref_{(\theta,n_{\theta})}[1]=& \left\lceil \frac{2d-2n_{\theta}X_{\theta,T}-(n_{\theta}+2)X_{\theta,R}}{X_{\theta,R}}\right.
	\\
	&\left.-\frac{4l_{1}\tan{\arcsin{\frac{n_{1}\sin{\theta}} {n_{3}}}}}{X_{\theta,R}}
	\right\rceil
	\label{ref1}
	\end{aligned},
\end{equation}

\begin{equation}
\begin{aligned}
	ref_{(\theta,n_{\theta})}[y_{(\theta,n_{\theta})}]&= \left\lfloor 
	\frac{2d+2L-2n_{\theta}X_{\theta,T}-(n_{\theta}+2)X_{\theta,R}}{X_{\theta,R}}\right.\\
	&\left.-\frac{4l_{1}\tan{\arcsin{\frac{n_{1}\sin{\theta}} {n_{3}}}}}{X_{\theta,R}}
	\right\rfloor
	\label{equ:refth}
\end{aligned},
\end{equation}

\begin{equation}
	y_{(\theta,n_{\theta})}= ref_{(\theta,n_{\theta})}[y_{(\theta,n_{\theta})}]-ref_{(\theta,n_{\theta})}[1]+1
	\label{equ:ytheta},
\end{equation}

Based on (\ref{ref1}) and (\ref{equ:ytheta}), all these possible reflection times constitute a matrix  {$Ref_{(\theta,n_{\theta})}$}:

\begin{align}
	\textbf{Ref}_{(\theta,n_{\theta})}&=\left [
	\begin{matrix}
		ref_{(\theta,n_{\theta})}[1]\\
		ref_{(\theta,n_{\theta})}[2]\\
	    \vdots\\
		ref_{(\theta,n_{\theta})}[y_{(\theta,n_{\theta}})]\\
	\end{matrix}\right ] \label{equ_1h}
\end{align}

\subsection{Model of paths in border-less 3DWiNoCs}\label{A1A}
3DWiNoCs are formed by stacking multiple single-layer NoCs.
When $\theta$ or the number of NoC layers is large, the signal can not reach the boundary of the NoCs.
This case is defined  as border-less model in this paper.

The paths with the same number of reflect and refract rays are classified as a specific class of path with the same channel gain.
Therefore, the total channel gain of a class can be calculated by multiplying the individual channel gain and the number of the path in the class.
The number of paths in such a specific class (consisting of $n_{\theta}$ refractions and $m_{\theta}$ reflections) is  $Path_{n_{\theta},m_{\theta}}$.

\begin{figure}[htbp]
\center
\centering\includegraphics[scale=0.5]{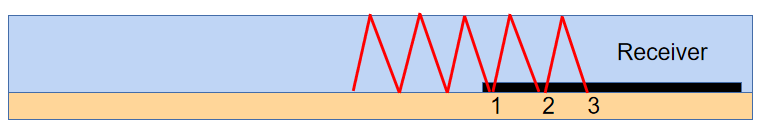}
\caption{Modeling an absorbed path.}
\label{fig_borderlessnoc}
\end{figure}

However, the energy of the signal reaching the receiver will be absorbed and all paths should reach the receiver only for the first time.
If a path has reached the receiver before the end point, the path is unpractical.
Fig. \ref{fig_borderlessnoc} show an example of the case in which the signal has been absorbed at point 1, and the paths of points 2 and 3 are unpractical.
The number of unpractical paths in each class is defined as $Path^{redundant}_{n_{\theta},m_{\theta}}$.

The number of received paths is defined as $Path^{total}_{n_{\theta},m_{\theta}}$:

\begin{equation}
	Path^{total}_{n_{\theta},m_{\theta}}=Path_{n_{\theta},m_{\theta}}- Path^{redundant}_{n_{\theta},m_{\theta}}
	\label{equ:antenna_rul3e}.
\end{equation}

The simulated transfer function of such a specific class (wthin $n_{\theta}$ times refraction and $m_{\theta}$ times reflection)
$G_{(n_{\theta},m_{\theta})}$ is:

\begin{equation}
\begin{aligned}
	&G_{(n_{\theta},m_{\theta})}= Path^{total}_{n_{\theta},m_{\theta}}e^{-3l_{3}\lambda_{3}-3l_{2}\lambda_{2}-3l_{1}\lambda_{1}}T^2_{1}T_{2}T_{3}T_{4}R_{4}\\
	&(T_{1}T_{4}T_{5}R_{4}e^{-2l_{3}\lambda_{3}-l_{2}\rho_{2}-l_{1}\lambda_{1}})^\frac{n_{\theta}-1}{2}(R_{3}R_{6}e^{-2l_{3}\lambda_{3}})^{m_{\theta}}
	\label{313}
	\end{aligned}.
\end{equation}

By adding all the possible specific class (based on determining all the possible $\theta$, all the possible refraction time $n_{\theta}$ and all the possible reflection $m-{\theta}$), the simulated transfer function \textit{H} is

\begin{equation}
	H=\int_{0}^{\theta_{bound}} 
	\sum_{n_{\theta}=J}^{tra_{\theta}[x_{\theta}]}
	\sum_{M_{\theta}=ref_{(\theta,n_{\theta})}[1]}^{ref_{(\theta,n_{\theta})}[y_{(\theta,n_{\theta})}]}
	G_{t}G_{r}
	G_{(n_{\theta},m_{\theta})} 
	d\theta ,
	\label{equ:ante4312le6}
\end{equation}
where $G_{t}$ and $G_{r}$ are the simulated gains for the transmitter and receiver, respectively.

\subsection{Model of paths in boundary-constrained 3DWiNoCs}\label{A11A}

In practice, boundaries exist in 3DWiNoCs, which makes the paths cross out of a boundary not be received. 
When the transmitter or receiver is close to the boundary, many paths will be unpractical.
Thus,the factor of boundary must be considered in this case.

\begin{figure}[htbp]
\center
\centering\includegraphics[scale=0.6]{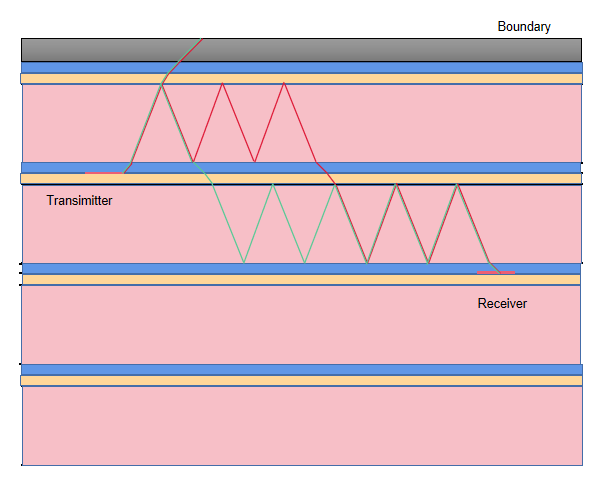}
\caption{Model of paths in boundary-constrained 3DWiNoCs.}
\label{fig_systemamodel1}
\end{figure}

$J_{bound}$ is used to defined the number of layers between the transmitter and the nearest boundary.
When the the difference between the upward refraction and downward refraction is  higher than $J_{bound}$ (see Fig. 6), the path leaves the chip through a boundary.
Thus the number of total paths in boundary-constrained model can be obtained as $Path^{bound}_{n_{\theta},m_{\theta}}$:

\begin{equation}
	Path'^{total}_{n_{\theta},m_{\theta}}=Path_{n_{\theta},m_{\theta}}- Path^{redundant}_{n_{\theta},m_{\theta}}-Path^{bound}_{n_{\theta},m_{\theta}}
	\label{equ:antenna_r}.
\end{equation}

The simulated transfer function of a path class
$G'_{(n_{\theta},m_{\theta})}$ is

\begin{equation}
\begin{aligned}
	G'_{(n_{\theta},m_{\theta})}= Path'^{total}_{n_{\theta},m_{\theta}}e^{-3l_{3}\lambda_{3}-3l_{2}\lambda_{2}-3l_{1}\lambda_{1}}T^2_{1}T_{2}T_{3}T_{4}R_{4}\\
	(T_{1}T_{4}T_{5}R_{4}e^{-2l_{3}\lambda_{3}-l_{2}\lambda_{2}-l_{1}\lambda_{1}})^\frac{n_{\theta}-1}{2}(R_{3}R_{6}e^{-2l_{3}\lambda_{3}})^{m_{\theta}}
	\label{equ:antenna_rule00}
\end{aligned},
\end{equation}
and the simulated transfer function \textit{H} is
\begin{equation}
\begin{aligned}
	H&=\int_{0}^{\theta_{bound}} 
	\sum_{n_{\theta}=J}^{tra_{\theta}[x_{\theta}]}
	\sum_{M_{\theta}=ref_{(\theta,n_{\theta})}[1]}^{ref_{(\theta,n_{\theta})}[y_{(\theta,n_{\theta})}]}
	G_{t}G_{r}
	G'_{(n_{\theta},m_{\theta})} 
	d\theta
	\label{equ:a0ntenna_rule}
	\end{aligned}.
\end{equation}

\subsection{Complexity analysis of the two models}

In practical calculation,continuous integration in (16) and (19) needs to be transformed into sampling and discrete summation.
The outermost loop is to calculate an approximated integration via summation over discrete angles $\theta$=$\frac{1}{r}\theta_{bound}$,$\frac{2}{r}\theta_{bound}$,$\cdots$,$\theta_{bound}$, and the
 sampling time is $r$ (larger $r$ leads to higher calculation precision).

Although boundary-constrained model is practical, it has to implement more calculation loops to exclude the paths transmitting out of the boundary ($Path_{n_{\theta},m_{\theta}}^{bound}$ in (17)),  compared with boundary-less model.
The additional loops cause by paths transmitting out of the boundary is given as (\ref{complexity}). 

\begin{equation}
\begin{aligned}
&Loop_{BC}-Loop_{BL}\\
=&\sum_{\theta=0}^{\theta_{bound}}
\frac{\alpha}{2}(J_{bound}+\beta)(\beta-J_{bound}+1)\\
=&\sum_{\theta=0}^{\theta_{bound}}\frac{\alpha}{2}(-J_{bound}^2+J_{bound}+\beta^2+\beta),
\label{complexity}
\end{aligned}
\end{equation}
where $\alpha$ and $\beta$ are variables associated with $\theta$ and $d$, $L$, thickness and  refractive index of materials, $J_{bound}$ is the number of layer between transmitter and boundary, and $\theta$=$\frac{1}{r}\theta_{bound}$,$\frac{2}{r}\theta_{bound}$,$\cdots$,$\theta_{bound}$.
The proof of (\ref{complexity}) is shown in \textbf{Appendix A}.

Equ. (\ref{complexity}) shows that the boundary-less model is much simpler than the boundary-constrained one.
If $J_{bound}$ increases, the difference between the two models becomes minor (the difference is proportional to squared of $J_{bound}$).
When $J_{bound}$ approaches to infinity, the complexity of boundary-less model equals to boundary-constrained model.

\section{Calculation of Path Number}

\subsection{Calculation of $Path_{n_{\theta},m_{\theta}}$}

The number of path in a specific class is determined by the combinatorial number of reflect and refract rays. 

To start with the calculation, the combinatorial number of refract rays should be obtained.
Because the upward refraction and downward refraction effects are both exist in the path, refract rays can be classified into upward groups and downward groups to facilitate the classification of these refraction rays.
After $n_{\theta}$ refract times, the transmitted signal finally reaches the receiver's antenna, which is $J$ layers away from the transmitter.
The refraction process is selection into a vector, where each elements is 1 or -1 and the length is  $n_{\theta}$.
The summation of all entries in this vector is $-J$.
These factors can be formulated in (\ref{equ:calc})

\begin{equation}  
\left\{  
             \begin{array}{lr}  
             \textit{P}=[\pm1,\pm1...\pm1] &  \\ 
             Length(\textit{P})=n_{\theta} &  \\ 
         \sum_{i=J}^{n_{\theta}}\textit{P}[i]=-J
              &    
             \end{array}.  
\right.  
\label{equ:calc}
\end{equation}

\begin{figure}[htbp]
\center
\centering\includegraphics[scale=0.45]{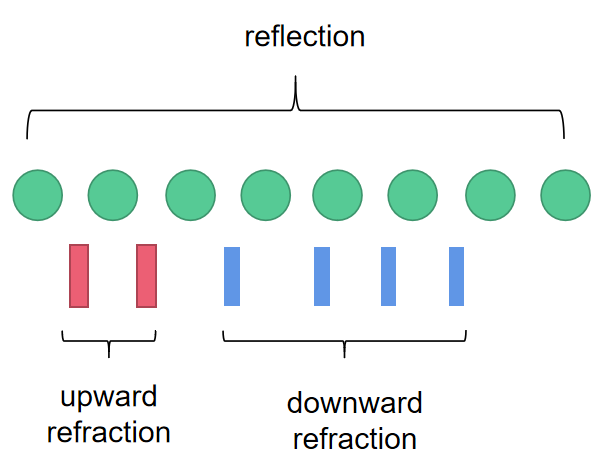}
\caption{Number of combinations}
\label{simple}
\end{figure}

 We use Fig. \ref{simple} to describe the   permutation and combination problems, where the refract rays are marked as rectangle (upward refraction is blue and downward refraction is red) and the reflect rays are marked as circle.

In determining the order of rectangle, the number of refraction (the rectangle in Fig. (\ref{simple})) combinations can be obtained as $Com_{T}({n_{\theta}},J)$ in (\ref{equ:Com}).

\begin{equation}
	Com_{T}({n_{\theta}},J)=C^{\frac{n_{\theta}-J}{2}}_{n_{\theta}}
	\label{equ:Com}.
\end{equation}

In a specific class, calculating the number of paths in the class can be formulated as a partitioning problem, where the reflect rays are randomly put on sides of the refract rays.
After the order of rectangle is determined, the trangles should be randomly put into the gaps between rectangles.
Thus, the number of all the possible schemes equals the summation of path.
The number of the paths in a specific class equals to the multiplication of two combinatorial number (shown in (23)).

  \begin{equation}
   Path_{n_{\theta},m_{\theta}}=Com_{T}({n_{\theta}},J)C^{m_{\theta}}_{m_{\theta}+n_{\theta}}.
	\label{equ:anten4na_rule}
   \end{equation}

\subsection{Calculation of $Path^{total}_{n_{\theta},m_{\theta}}$}

The combination of refraction and reflection based on the ergodicity of all possibilities of paths.
However, some paths do not exist because the antenna absorbs all the energy when the path first reaches to the receiver (see Fig. 4).
To reduce the error caused by unpractical paths, unpractical paths must excluded.

To  start  with  the total calculation of paths, one must obtain the lowest time of refraction $n'_{\theta}$ of all the signals that can reach the receiver layer.
Since $n'_{\theta}$ is the minimum number of refraction ($n'_{\theta}=J$ means the path reaches the receiver at first time and have never been absorbed by receiver), there are no unpractical paths and these paths has $n'_{\theta}$ refract trays and $m_{\theta}$ reflect rays. 

The number of unpractical path is direct proportion with the number of refraction.
However, for each class with a certain refraction, we can exclude the number of unpractical paths by subtracting the number of paths of specific class in adjacent refraction numbers.
Therefore, do the loop for each $n'_{\theta}$  until reaching the certain number of refraction, and then we can get the number of path in such specific classes.

\begin{figure}[htbp]
\center
\centering\includegraphics[scale=0.62]{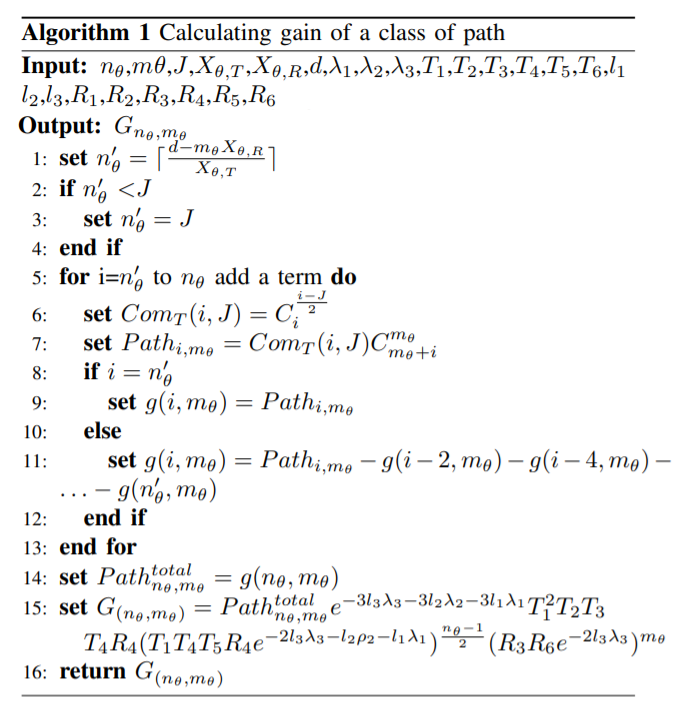}
\caption{Algorithm of gain caculation in boundary-less model.}
\label{fig_sddystemmodel1}
\end{figure}

\textbf{Algorithm 1} is proposed to calculate the channel gain of a path class (with the same number of refract and reflect rays) in boundary-less model.
The channel gain of all classes can be obtained by integration, where each gain of a class is calculated by \textbf{Algorithm 1}.

After getting the channel gain the a specific class (with the certain reflection and refraction time), we need to obtain every possible class and add them up to get the final channel gain $H_{boundaryless}$. Such a process need  to traverse all the possible transmission angle $\theta$, and then traverse all the possible refraction and reflection. Finally get the total channel gain $H_{boundaryless}$.
The process is shown in part of the proposed  \textbf{Algorithm 2}.

\begin{figure}[htbp]
\center
\centering\includegraphics[scale=0.61]{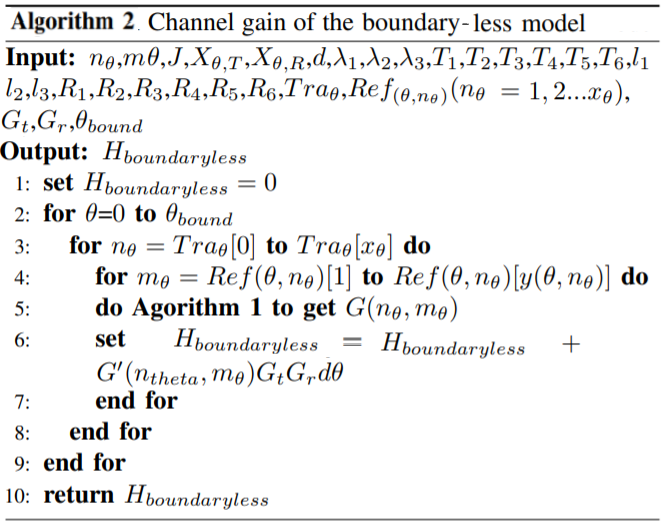}
\caption{Channel gain of the boundary-less model.}
\label{fig_systemqwmodel1}
\end{figure}

\subsection{Channel gain of boundary-constrained 3DWiNoC model}
When the transmitting angle is less than $\theta_{bound}$, the number of paths caused by boundary factors should be considered.

The calculation of number of paths in a specific class can be formulated as a combination problem of refract and reflect rays (see Fig. 7).
The refract rays are scheduled with different orders, and divided into two groups: upward and downward groups.
In this paper, the upward refraction group is defined as a number of 1, and the  downward refraction group is defined as a number of -1 (based on (21)).
The combination result of this process can be given as: $C^{\frac{n_{\theta}-J}{2}}_{n_{\theta}}$ -1, $J-C^{\frac{n_{\theta}-J}{2}}_{n_{\theta}}$, and $m_{\theta}$ (based on (23)).
The sum of any preceding elements in the array cannot exceed $J_{bound}$. Finally, the number of arrays that meets the requirements is the number of paths.

To start with the calculation, one must judge whether the boundary reduces the number of paths or not.
Some paths leave the boundary and, however, are still counted in $Path^{total}_{n_{\theta},m_{\theta}}$.
These paths should be excluded.
The number of possible refract rays beyond the boundary can be obtained by calculating the number of combinations of these refract rays.
In this process, redundancies exists because the next refraction contains part of the previous refraction.
Then, subtract the repeated parts by turning through the loop, and then add up all the traversal values to obtain $Path'^{total}_{n_{\theta},m_{\theta}}$.
After that, the gain of these paths can be obtained as ($G'(n_\theta,m_\theta)$), which is the gain for a specific class of paths (with the same number of refract and reflect rays). 

\begin{figure}[htbp]
\center
\centering\includegraphics[scale=0.61]{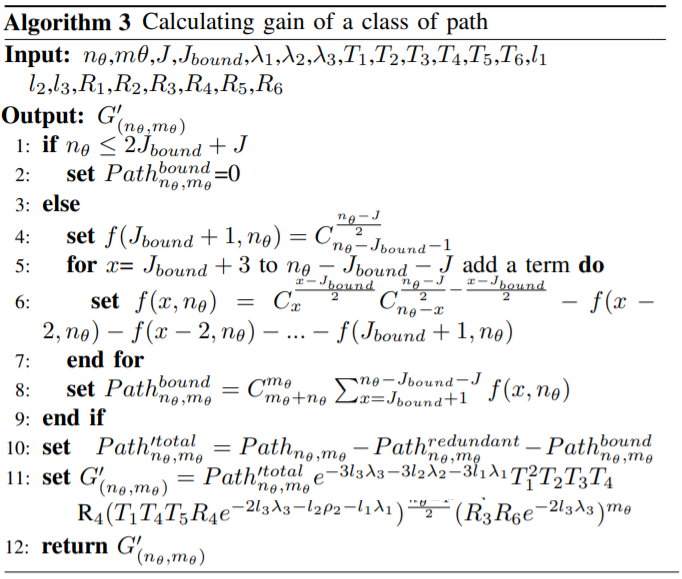}
\caption{Algorithm of gain caculation in boundary-constrained model.}
\label{fig_systedasdammodel1}
\end{figure}

After looping over all the classes and get each $G'(n_\theta,m_\theta)$, $H_{boundary}$ (channel gain of the boundary-constrained model) can be obtained by summation (see \textbf{Algorithm 3}).
All classes of paths can be looped by integrating $\theta$, and the path gain of each class is calculated by \textbf{Algorithm 1}.
Finally, the total channel gain $H_{boundary}$ of boundary-constrained model can be calculated by adding up all the possible path gain in all transmission angle $\theta$.
The process is shown in part of the proposed \textbf{Algorithm 4}.

\begin{figure}[htbp]
\center
\centering\includegraphics[scale=0.51]{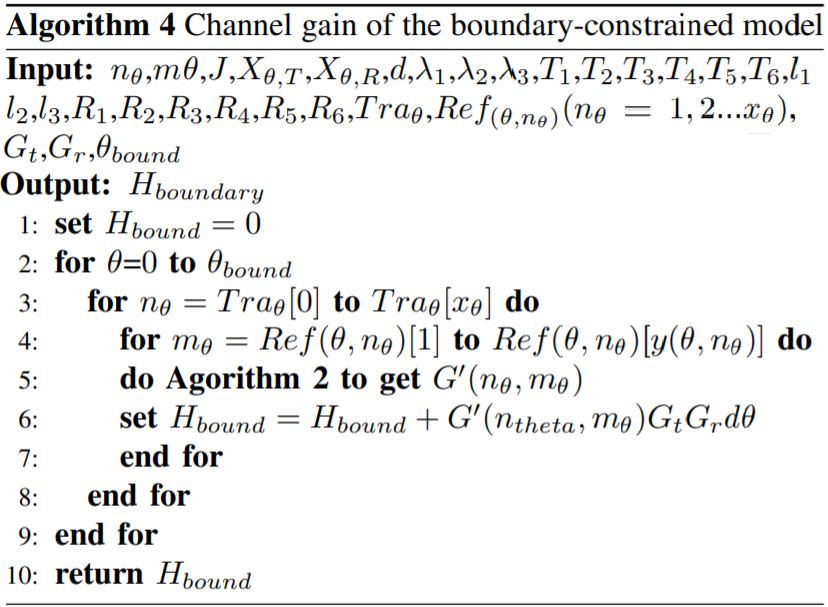}
\caption{Calculation of the channel gain for the boundary-constrained model.}
\label{fig_systemxcxzmodel1}
\end{figure}

\subsection{Approximate calculation of number of paths}
To finish the calculation of channel gain, the number of loops in the algorithm may exceed $10^{30}$ due to the enormous number of all the path classes.
As a result, the complexity of the algorithm is high because of the exponential increase of loops with $d$.
Therefore, some paths with minimal gain or high transmission delay could be ignored.
If these paths are ignored, the approximated channel gain will exhibit a minor difference compared to the actual one.

\begin{theorem} 
Consider: (1) The refractive indexes of three different materials layers in 3DWiNoC are $n_{1}$, $n_{2}$, and $n_{3}$, respectively. (2) The three material layers, i.e., Material 1, 2 and 3 are located at one single NoC in 3DWiNoC from top to bottom. (3) The rays refract through material 1 and 2. (4) The rays only reflect on material 3.

The ratio between the gain of one time of refraction and reflection is $ \frac{(n_3-n_1)^2(n_3-n_2)^2(n_2-n_1)^2\left(1-\frac{(n_3-n_2)^2}{(n_3+n_2)^2}\right)^2}{(n_3+n_1)^2(n_2+n_1)^2(n_3-n_2)^2}e^{\frac{2\lambda}{n_{3}}l_{3}-\frac{\lambda}{n_{2}}l_{2}-\frac{\lambda}{n_{1}}l_{1}}$. The channel gain of the reflect ray is much higher than the refract ray.

\end{theorem}

\emph{Proof 1 :} See Appendix B.

\emph{Remark 1:} Based on \textit{Theorem 1}, the gain of reflect rays is much higher than that of refract rays (about $10^{10}$ based on Fig. \ref{fig_gainRR}).
When the materials are $Si_3N_4$, $SiO_2$, and $Si$, the coefficient (the relation between refraction and reflection) is approximately $7.2\times10^{8}$.
Therefore, the gain of a path with excessive refract rays is small.
Fig. 14 shows the gap between the gains of reflect and refract rays. 

\begin{theorem} 
Consider: (1) The  coherence time of the channel is $t_c$. (2) The minimum horizontal displacement path is within the transmission angle $\theta_{bound}$. (3) The speed of light in a vacuum is $v$. (4) The path has $n_\theta$ transmissions and $m_\theta$ reflections.

The path transmitted with a angle $\theta$ less than $\theta_t=arctan\left(\frac{n_3(2m_{bound}+n_{bound}+1)l_3tan\theta_{bound}}{n_3(2m_\theta+n_\theta+1)l_3+vt_{c}tan\theta_{bound}}\right)$ cannot reach the receiver in coherent time.
\end{theorem}

 \emph{Proof 2 :} See Appendix C.

\textit{Remark 2:} Based on \textit{Theorem 2}, the signal with a sufficiently small transmission angle cannot reach the receiver within coherent time and may cause inter-symbol-interference.

\begin{figure}[htbp]
\center
\centering\includegraphics[scale=0.57]{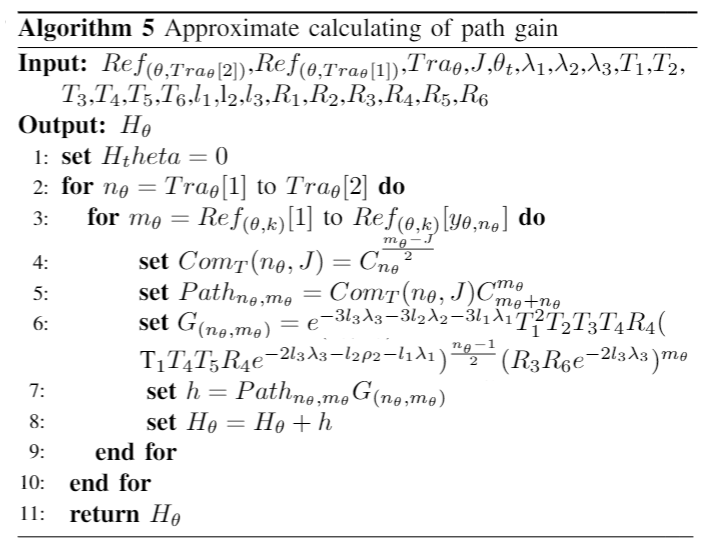}
\caption{Approximation algorithm of gain caculation.}
\label{fig_systemmobnvdel1}
\end{figure}

According to \emph{Theorem 1}, the approximation algorithm only calculates the path with the least number of refract rays with an error rate less than $1\%$ (see Fig. \ref{fig_gainRR}).
According to \emph{Theorem 2}, the approximation algorithm only calculates the path that arrived before coherent time.

Based on the aforementioned analysis, the \textbf{Algorithm 5} is able to simplify the sampling and the refraction determining processes.
The gain of the signal reaching the receiver at each angle is calculated as $H_\theta$ (see \textbf{Algorithm 3}).

\section{Numerical Results}
In this paper, the materials of stacking layers in 3DWiNoC [9] are selected as $Si_3N_4$, $SiO_2$, and $Si$.
We set sampling points as 10 $\theta$s in calculating integration.
As long antenna cannot be used inside the small NoCs and high frequency causes high attenuation, the signal frequency used is 1 $THz$.
\subsection{Analysis of path gain}
Fig. 13 shows the channel amplitudes of boundary-constrained and boundary-less 3DWiNoCs in different traveled layers ($J$).
As $J$ increases, the number of layers that signal travels through increase as well as the attenuation.
The channel gain decreases by -50 to -70 dB for each layer traveled, which is consistent with the results in [7].
Due to the existence of  boundaries, part of the paths refract 3DWiNoCs out and unable to be received, which makes the channel gain of boundary-constrained 3DWiNoCs lower than that of boundary-less ones.
However, this part of the path are excessively refracted, and thus the gain is minor (see Fig. 16).
Thus, the difference between the two model is very small (the difference further decreases as $J$ increases).

\begin{figure}[htbp]
\center
\centering\includegraphics[scale=0.45]{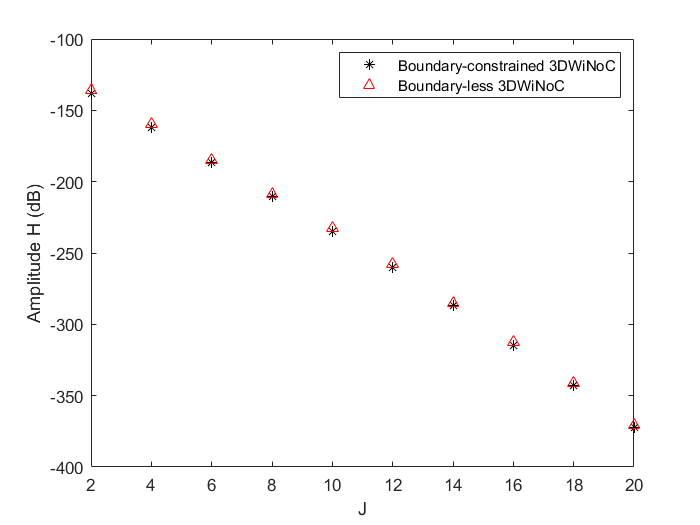}
\caption{Path gains of boundary-less and boundary-constrained 3DWiNoCs.}
\label{fig_systemmgfgodel1}
\end{figure}

Fig. 14 shows that the difference ($H_{boundaryless}-H_{boundary}$) between the two model is lower than $10^{-5}$, which means the boundary-constrained model  can be replaced by the computational efficient  boundary-less model with acceptable accuracy.

\begin{figure}[htbp]
\center
\centering\includegraphics[scale=0.38]{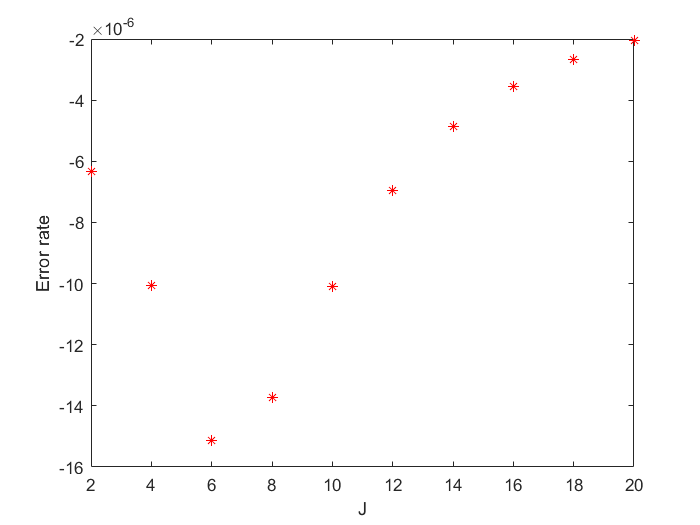}
\caption{The error rate of replacing boundary-constrained model by boundary-less model.}
\label{fig_systemeqwemodel1}
\end{figure}

Fig. 15 shows the channel gain of boundary-less 3DWiNoC at different horizontal displacements ($d$).
Setting the traveled layer ($J$) as 2, the channel gain decreases as horizontal  displacement increases.
When horizontal communication displacement increases, the number of reflect rays contained in the path increases (because the channel containing too many refract rays is ignored in the approximation algorithm based on \emph{Theorem 1}).
As seen in Fig. 16, the reflected signal attenuation is low, so the horizontal displacement has much less impact on the channel gain compared with the number of traveled layers ($J$).
\begin{figure}[htbp]
\center
\centering\includegraphics[scale=0.38]{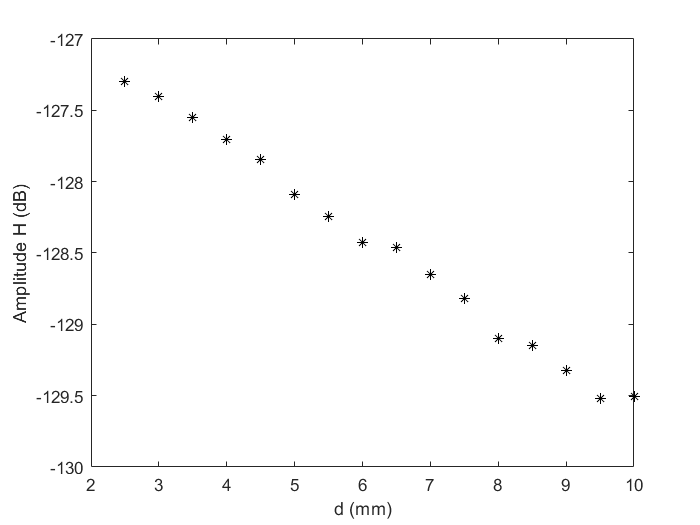}
\caption{Channel gains at different horizontal  displacements ($d$).}
\label{fig_systxcfsemmodel1}
\end{figure}

\subsection{Analysis of approximation calculation}
\begin{figure}[htbp]
\center
\centering\includegraphics[scale=0.38]{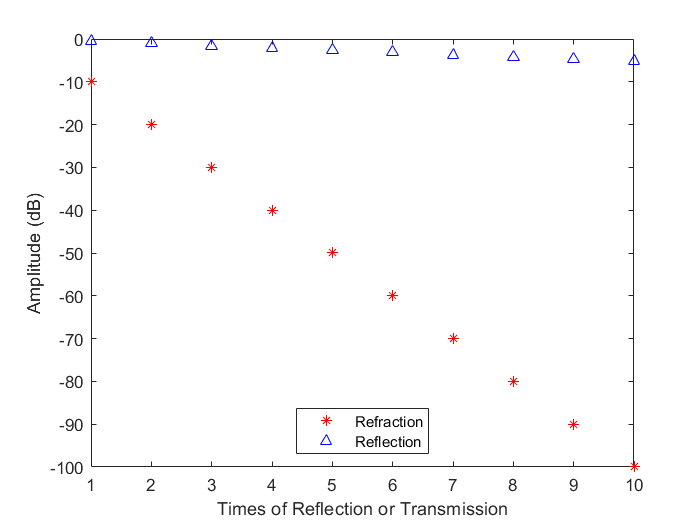}
\caption{Gains of refract and reflect rays.}
\label{fig_gainRR}
\end{figure}

More than $10^{30}$ paths are involved in the completed channel model, which makes the number of loops of the algorithm large.
The gains of many of these paths, however, are small, so that they can be ignored.
Numerical results show that the path gain decreases sharply with increasing number of refract rays, while the effect of reflections on the path gain is relatively low (see Fig. \ref{fig_gainRR}). 

\begin{figure}[htbp]
\center
\centering\includegraphics[scale=0.38]{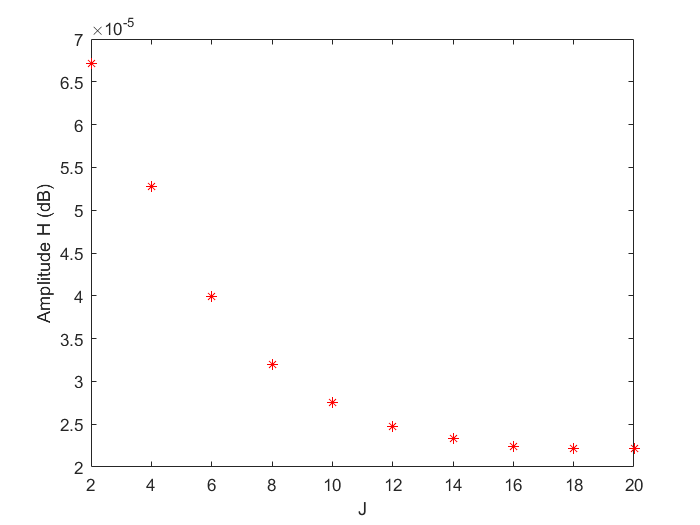}
\caption{Difference between approximate and realistic gains.}
\label{fig_diffince}
\end{figure}

To obtain the channel gain, the traversal of the channel in the complete or realistic model needs four loops, while the approximate algorithm only needs three loops.

Fig. \ref{fig_diffince} shows the gap between the approximate and gains of the accurate channel model, and it is approximately $5\times10^{-5}$ dB (decreasing with increasing $J$).
Compared with the channel amplitude, the difference is approximately one-thousandth of the aforementioned value.

\section{Conclusions}

In this paper, boundary-constrained and boundary-less models are proposed to characterize the separate cases in which a translator is far from and close to the boundary of 3DWiNoCs, respectively.
Compared with single-layer NoC channel model, the proposed models can be used to comprehensively characterize the complex reflections and refractions in NoC stacking 3D situation.
The paths are classified by the number of reflect and refract rays, and the number of each respective path class can be calculated. Numerical results show that the gain attenuation of the signal travel through a layer is about -60$dB$  and the difference between the two models (boundary-less model and boundary-constrained model) is lower than $10^{-5}$.
In addition, an approximation algorithm is proposed to obtain the channel gain with lower complexity.

\section{Acknowledgement}
This work is supported in part by Guangdong Province Basic and Applied Basic Research Fund (Grant No. 2019A1515111086), in part by the Fundamental Research Funds for the Central Universities (Grant No. FRF-BD-20-11A), in part by the National Natural Science Foundation of China (Grant No. 62001019), and in part by Beihang Hefei Science City Project (Grant No. 01-0004-01) from Beihang University Hefei Innovation Research Institute.

\begin{appendices}
\section{Proof of computation complexity}
The complexity of the model mainly depends on the number of computation loops.
In practical calculation, integration needs to be transformed into discrete sampling, which changes the continuous calculation into discrete calculation to form the outermost loop.
Integral over angles can be transformed into sum over some discrete angles to determine the number of refract and reflect rays.
Therefore, the sampling number of outermost loop is $r$ with sample points $\theta$=$\frac{1}{r}\theta_{bound}$,$\frac{2}{r}\theta_{bound}$,$\cdots$,$\theta_{bound}$.
After the number of refract and reflect rays is determined, the gain calculation of boundary-constrained model needs computation of a conbinatorial number and has to traverse the paths that travel out of the boundary.
Thus, the number of computation loops for boundary-less model and boundary-constrained model are $Loop_BL$ and $Loop_BC$, respectively.

\begin{equation}
Loop_{BC}=\sum_{\theta=0}^{\theta_{bound}} 
	\sum_{n_{\theta}=J}^{tra_{\theta}[x_{\theta}]}
	\sum_{M_{\theta}=ref_{(\theta,n_{\theta})}[1]}^{ref_{(\theta,n_{\theta})}[y_{(\theta,n_{\theta})}]}
	(2n_{\theta}-J_{bound})
\end{equation}

\begin{equation}
Loop_{BL}=\sum_{\theta=0}^{\theta_{bound}} 
	\sum_{n_{\theta}=J}^{tra_{\theta}[x_{\theta}]}
	\sum_{M_{\theta}=ref_{(\theta,n_{\theta})}[1]}^{ref_{(\theta,n_{\theta})}[y_{(\theta,n_{\theta})}]}n_\theta
\end{equation}

The difference of the number of loops between the two model is:
\begin{small}
\begin{equation}
\begin{aligned}
&Loop_{BC}-Loop_{BL}=\sum_{\theta=0}^{\theta_{bound}} 
\sum_{n_{\theta}=J}^{tra_{\theta}[x_{\theta}]}
\sum_{M_{\theta}=ref_{(\theta,n_{\theta})}[1]}^{ref_{(\theta,n_{\theta})}[y_{(\theta,n_{\theta})}]}(n_{\theta}-\\
&J_{bound})\\
&=\sum_{\theta=J}^{\theta_{bound}} 
\sum_{n_{\theta}=J}^{tra_{\theta}[x_{\theta}]}
({ref_{(\theta,n_{\theta})}[y_{(\theta,n_{\theta})}]}-{ref_{(\theta,n_{\theta})}[1]}+\\
&1)(n_{\theta}-J_{bound})
\end{aligned}
\end{equation}
\end{small}

Referring to (11) and (12), (26) can be rewritten as:
\begin{small}
\begin{equation}
\begin{aligned}
&Loop_{BC}-Loop_{BL}=\sum_{\theta=0}^{\theta_{bound}} 
\sum_{n_{\theta}=J}^{tra_{\theta}[x_{\theta}]}
\left\lceil\frac{L}{l_2tan\theta}+1\right\rceil
(n_{\theta}-\\
&J_{bound})\\
&=\frac{1}{2}\sum_{\theta=J}^{\theta_{bound}}
\left\lceil\frac{L}{l_2tan\theta}+1\right\rceil
(J_{bound}+tra_{\theta}[x_{\theta}](tra_{\theta}[x_{\theta}]-\\
&J_{bound}+1))
\end{aligned}
\end{equation}
\end{small}

Referring to (7) and (8), (27) can be written as:

\begin{small}
\begin{equation}
\begin{aligned}
&Loop_{BC}-Loop_{BL}=\frac{1}{2}\sum_{\theta=0}^{\theta_{bound}}
\left\lceil\frac{L}{l_2tan\theta}+1\right\rceil \times \\
&(J_{bound}+tra_{\theta}[x_{\theta}](tra_{\theta}[x_{\theta}]-J_{bound}+1)\\
&=\frac{1}{2}\sum_{\theta=J}^{\theta_{bound}}
\left\lceil\frac{L}{l_2tan\theta}+1\right\rceil \times\\
&\left(J_{bound}+
\left\lfloor \frac{2(d+L-2l_{1}\tan{\arcsin{\frac{n_{1}\sin{\theta}} {n_{3}}}}-X_{\theta,R})}{2X_{\theta,T}+X_{\theta,R}} \right\rfloor\right)\times\\
&\left (\left\lfloor \frac{2(d+L-2l_{1}\tan{\arcsin{\frac{n_{1}\sin{\theta}} {n_{3}}}}-X_{\theta,R})}{2X_{\theta,T}+X_{\theta,R}} \right\rfloor-J_{bound}+1\right )
\end{aligned}
\end{equation}
\end{small}

Referring to (3) and (4), (28) can be written as:

\begin{small}
\begin{equation}
\begin{aligned}
&Loop_{BC}-Loop_{BL}=\frac{1}{2}\sum_{\theta=0}^{\theta_{bound}}\left\lceil\frac{L}{l_2tan\theta}+1\right\rceil\times(J_{bound}+\\
&\left\lfloor \frac{2(d+L-2l_{1}\tan{\arcsin{\frac{n_{1}\sin{\theta}} {n_{3}}}}-2l_{2}\tan{\theta})}{2l_{1}\tan{\arcsin{\frac{n_{1}\sin{\theta}} {n_{3}}}}+2l_{2}\tan{\arcsin{\frac{n_{2}\sin{\theta}} {n_{3}}}}+2l_{2}\tan{\theta}} \right\rfloor)\times\\
&\left(\left\lfloor \frac{2(d+L-2l_{1}\tan{\arcsin{\frac{n_{1}\sin{\theta}} {n_{3}}}}-2l_{2}\tan{\theta})}{2l_{1}\tan{\arcsin{\frac{n_{1}\sin{\theta}} {n_{3}}}}+2l_{2}\tan{\arcsin{\frac{n_{2}\sin{\theta}} {n_{3}}}}+2l_{2}\tan{\theta}} \right\rfloor-\right.\\
&\left.J_{bound}+1\right)
\end{aligned}
\end{equation}
\end{small}

Using intermediate variables to simplify the (29), the formula can be rewritten as :
\begin{equation}
\begin{aligned}
&Loop_{BC}-Loop_{BL}=\sum_{\theta=0}^{\theta_{bound}}
\frac{\alpha}{2}(J_{bound}+\beta)(\beta-J_{bound}+1)\\
&=\frac{1}{2}\sum_{\theta=0}^{\theta_{bound}}\alpha(-J_{bound}^2+J_{bound}+\beta^2+\beta),
\end{aligned}
\end{equation}
where the intermediate variables $\alpha$ and $\beta$ are functions depended on $\theta$, $L$, $d$, the thickness and refractive index of materials:
\begin{equation}
\begin{aligned}
\alpha(\theta,L,l_2)=\left\lceil\frac{L}{l_2tan\theta}+1\right\rceil
\end{aligned}
\end{equation}

\begin{small}
\begin{equation}
\begin{aligned}
&\beta(\theta,L,d,l_1,l_2,n_1,n_2,n_3)=\\
&\left\lfloor \frac{(d+L-2l_{1}\tan{\arcsin{\frac{n_{1}\sin{\theta}} {n_{3}}}}-2l_{2}\tan{\theta})}{l_{1}\tan{\arcsin{\frac{n_{1}\sin{\theta}} {n_{3}}}}+l_{2}\tan{\arcsin{\frac{n_{2}\sin{\theta}} {n_{3}}}}+l_{2}\tan{\theta}} \right\rfloor
\end{aligned}
\end{equation}
\end{small}

Therefore, the difference of the computation complexity between boundary-less model and boundary-constrained one is negative correlation to  $J_{bound}^2$.

\section{Proof of Theorem 1}
The gain of a refraction is  $G_{tra}=T_{3}e^{-\lambda_{2}l_{2}}T_{2}e^{-\lambda_{1}l_{1}}T_{1}$ and the gain of a reflection is $G_{ref}=R_{3}R_{6}e^{-2\lambda_{3}l_{3}}$.
Their ratio can be obtained as

\begin{equation}
\begin{aligned}
G_{ratio}&=\frac{T_{3}e^{-\lambda_{2}l_{2}}T_{2}e^{-\lambda_{1}l_{1}}T_{1}}{R_{3}R_{6}e^{-2\lambda_{3}l_{3}}}  \\
&= \frac{T_{3}T_{2}T_{1}}{R_{3}R_{6}}e^{2\lambda_{3}l_{3}-\lambda_{2}l_{2}-\lambda_{1}l_{1}} \\
&= \frac{T_{3}T_{2}T_{1}}{(1-T_{3})(1-T_{6})}e^{2\lambda_{3}l_{3}-\lambda_{2}l_{2}-\lambda_{1}l_{1}}
\end{aligned}
\end{equation}
which can be easily derived by (1).

\section{Proof of Theorem 2}
The displacement traveled through by the path is mainly in $Si$.
Thus, based on (6), the approximated minimum reaching time of each $\theta$ is
\begin{equation}
t_{\theta}=\left(\frac{(2m_{\theta}+1)l_{3}}{tan\theta}+\frac{+n_{\theta}l_3}{tan\theta}\right)\frac{n_3}{v}.
\end{equation}

Based on (17), The time of arrival of the minimum path satisfies
\begin{equation}
t_{min}=\left(\frac{(2m_{\theta_{bound}}+1)l_{3}}{tan\theta_{bound}}+\frac{n_{\theta_{bound}}l_3}{tan\theta_{bound}}\right)\frac{n_3}{v}.
\end{equation}

The path within transmission angle $\theta_t$ to the receiver is exactly at $t_{min}+t_{c}$, so the $\theta_t$ satisfies

\begin{equation}
\theta_t=arctan\left(\frac{n_3(2m_{bound}+n_{bound}+1)l_3tan\theta_{bound}}{n_3(2m_\theta+n_\theta+1)l_3+vt_{c}tan\theta_{bound}}\right).
\end{equation}
\end{appendices}
\balance
\bibliographystyle{IEEEtran}
\bibliography{ref}

\begin{thebibliography}{1}
\providecommand{\url}[1]{#1}
\csname url@samestyle\endcsname
\providecommand{\newblock}{\relax}
\providecommand{\bibinfo}[2]{#2}
\providecommand{\BIBentrySTDinterwordspacing}{\spaceskip=0pt\relax}
\providecommand{\BIBentryALTinterwordstretchfactor}{4}
\providecommand{\BIBentryALTinterwordspacing}{\spaceskip=\fontdimen2\font plus
\BIBentryALTinterwordstretchfactor\fontdimen3\font minus
  \fontdimen4\font\relax}
\providecommand{\BIBforeignlanguage}[2]{{%
\expandafter\ifx\csname l@#1\endcsname\relax
\typeout{** WARNING: IEEEtran.bst: No hyphenation pattern has been}%
\typeout{** loaded for the language `#1'. Using the pattern for}%
\typeout{** the default language instead.}%
\else
\language=\csname l@#1\endcsname
\fi
#2}}
\providecommand{\BIBdecl}{\relax}
\BIBdecl

\bibitem{5306555}
A.~{Horibe} and F.~{Yamada}, ``Advanced 3d chip stack process for thin dies
  with fine pitch bumps using pre-applied inter chip fill,'' in \emph{2009 IEEE
  International Conference on 3D System Integration}, 2009, pp. 1--4.

\bibitem{4550025}
J.~U. {Knickerbocker}, P.~S. {Andry}, B.~{Dang}, R.~R. {Horton}, C.~S. {Patel},
  R.~J. {Polastre}, K.~{Sakuma}, E.~S. {Sprogis}, C.~K. {Tsang}, B.~C. {Webb},
  and S.~L. {Wright}, ``3d silicon integration,'' in \emph{2008 58th Electronic
  Components and Technology Conference}, 2008, pp. 538--543.

\bibitem{6991560}
D.~{DiTomaso}, A.~{Kodi}, D.~{Matolak}, S.~{Kaya}, S.~{Laha}, and W.~{Rayess},
  ``A-winoc: Adaptive wireless network-on-chip architecture for chip
  multiprocessors,'' \emph{IEEE Transactions on Parallel and Distributed
  Systems}, vol.~26, no.~12, pp. 3289--3302, Dec 2015.

\bibitem{959649}
{Kihong Kim}, W.~{Bomstad}, and K.~O. {Kenneth}, ``A plane wave model approach
  to understanding propagation in an intra-chip communication system,'' in
  \emph{IEEE Antennas and Propagation Society International Symposium. 2001
  Digest. Held in conjunction with: USNC/URSI National Radio Science Meeting
  (Cat. No.01CH37229)}, vol.~2, 2001, pp. 166--169 vol.2.

\bibitem{5551123}
A.~{Ganguly}, K.~{Chang}, S.~{Deb}, P.~P. {Pande}, B.~{Belzer}, and
  C.~{Teuscher}, ``Scalable hybrid wireless network-on-chip architectures for
  multicore systems,'' \emph{IEEE Transactions on Computers}, vol.~60, no.~10,
  pp. 1485--1502, 2011.

\bibitem{4339598}
Y.~P. {Zhang}, Z.~M. {Chen}, and M.~{Sun}, ``Propagation mechanisms of radio
  waves over intra-chip channels with integrated antennas: Frequency-domain
  measurements and time-domain analysis,'' \emph{IEEE Transactions on Antennas
  and Propagation}, vol.~55, no.~10, pp. 2900--2906, 2007.

\bibitem{6525613}
D.~W. {Matolak}, S.~{Kaya}, and A.~{Kodi}, ``Channel modeling for wireless
  networks-on-chips,'' \emph{IEEE Communications Magazine}, vol.~51, no.~6, pp.
  180--186, 2013.

\bibitem{8406954}
Y.~{Chen} and C.~{Han}, ``Channel modeling and analysis for wireless
  networks-on-chip communications in the millimeter wave and terahertz bands,''
  in \emph{IEEE INFOCOM 2018 - IEEE Conference on Computer Communications
  Workshops (INFOCOM WKSHPS)}, 2018, pp. 651--656.

\bibitem{5067340}
M.~{Sun}, Y.~P. {Zhang}, G.~X. {Zheng}, and W.~{Yin}, ``Performance of
  intra-chip wireless interconnect using on-chip antennas and uwb radios,''
  \emph{IEEE Transactions on Antennas and Propagation}, vol.~57, no.~9, pp.
  2756--2762, 2009.

\end{thebibliography}
\vspace{12pt}

\end{document}